\documentclass{article}

\usepackage{arxiv}

\usepackage[utf8]{inputenc} 
\usepackage[T1]{fontenc}    
\usepackage{hyperref}       
\usepackage{amsfonts}       
\usepackage{nicefrac}       
\usepackage{lipsum}		
\usepackage{doi}

\usepackage{amsmath}
\usepackage{graphicx}
\usepackage{microtype}
\usepackage{url}
\usepackage{float} 
\usepackage{makecell}
\usepackage{tabularx}
\usepackage{booktabs}
\usepackage{array}
\usepackage{graphicx}
\graphicspath{{figures/}}

\newcolumntype{Y}{>{\raggedright\arraybackslash}X}

\title{Tokenized but Illiquid? \\Evidence from Real-World Asset Markets}

\author{ \href{https://orcid.org/0000-0003-4424-3736}{\includegraphics[scale=0.06]{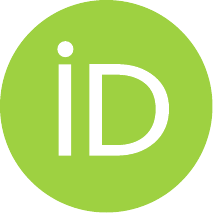}\hspace{1mm}Rischan Mafrur}\\
	School of Computer, Data and Mathematical Sciences\\
	Western Sydney University\\
	\texttt{R.Mafrur@westernsydney.edu.au} \\
}

\renewcommand{\shorttitle}{Tokenized but Illiquid? Evidence from Real-World Asset Markets}

\hypersetup{
pdftitle={A template for the arxiv style},
pdfsubject={q-bio.NC, q-bio.QM},
pdfauthor={David S.~Hippocampus, Elias D.~Striatum},
pdfkeywords={First keyword, Second keyword, More},
}

\begin{document}
\maketitle

\begin{abstract}
Real-world asset tokenization is often presented as a mechanism for improving the liquidity of traditionally illiquid assets. However, on-chain representation and secondary-market liquidity are distinct outcomes. This paper examines whether tokenized real-world assets exhibit meaningful observed liquidity and identifies the token characteristics associated with higher market activity. Using token-level data from RWA.xyz and supplemental contract-level observations from Etherscan, the study constructs an Ethereum-based monthly panel of non-stablecoin real-world assets across three prominent categories: U.S. Treasury-backed tokens, gold-backed commodity tokens, and private-credit-related tokens. Liquidity is measured using turnover, active addresses, and an active-month indicator. The empirical design combines descriptive statistics, non-parametric group tests, and exploratory panel regressions suited to short and sparse token histories. The results show substantial heterogeneity across asset categories. Gold-backed tokens exhibit broader holder bases and more persistent on-chain activity than many Treasury and private-credit-related products, while outstanding asset value alone does not reliably predict observed liquidity. The paper contributes to the literature by developing a clearer empirical measurement framework for real-world-asset liquidity and showing that tokenization and liquidity should be analyzed as distinct outcomes.
\end{abstract}


\section{Introduction}

Tokenization is increasingly presented as a way to modernize financial markets by making ownership records programmable, enabling fractional participation, and reducing operational frictions across issuance, settlement, and servicing processes \cite{WEF2025,BIS2024,IMF2025}. In this narrative, real-world assets such as government bonds, private credit, real estate, commodities, and fund shares can be represented as blockchain-based tokens and potentially become more accessible to a broader investor base. Yet the strongest practical claim surrounding tokenization is not simply that assets can be issued on-chain, but that tokenization can improve liquidity in markets that have historically been difficult to enter or exit.

That claim deserves closer scrutiny. Recent institutional work highlights the potential benefits of tokenization in transparency, composability, and post-trade efficiency, while also emphasizing persistent constraints related to fragmented regulation, limited interoperability, immature infrastructure, and weak secondary-market depth \cite{WEF2025,Nassr2025,IOSCO2025}. In other words, tokenization can change the technical form of ownership without necessarily producing active trading.

This distinction is especially important in the emerging market for tokenized real-world assets. Public dashboards now track a growing universe of non-stablecoin RWAs across multiple chains and platforms \cite{RWAxyzDashboard}. However, the distribution of tokenized value across categories says little by itself about the ease with which investors can trade, transfer, or exit those positions. A large tokenized market can remain illiquid if activity is concentrated in minting and redemption flows, if participation is narrow, or if tokens circulate only within limited institutional or operational settings.

The empirical literature remains narrower than the policy debate. Existing evidence is concentrated primarily in tokenized real estate. Swinkels studies 58 tokenized residential properties and shows that properties change hands roughly once per year on average \cite{Swinkels2023}. Kreppmeier et al.\ analyze 173 U.S. real-estate tokens and 238,433 blockchain transactions, finding broad ownership but limited diversification and important roles for crypto-market determinants in secondary activity \cite{Kreppmeier2023}. Laschinger et al.\ examine liquidity mechanisms in real-estate tokenization, while Cornelli shows that tokenized real estate can fill gaps in underserved local markets, although liquidity preservation during shocks depends on institutional buyback features \cite{Laschinger2024,Cornelli2025}. Taken together, these studies provide valuable evidence on tokenized real estate, but comparable cross-category empirical evidence for tokenized Treasuries, commodities, and private-credit-linked assets remains limited.

This paper addresses that gap. It asks two questions. First, do tokenized real-world assets exhibit meaningful observed liquidity once liquidity is measured directly rather than inferred from issuance growth? Second, which token characteristics are associated with greater observed liquidity across tokens and over time? To answer these questions, the paper constructs an Ethereum-based token-month panel using RWA.xyz as the primary data source, with Etherscan used selectively for contract-level validation and contextual cross-checking. Note that this study is limited to assets on the Ethereum network.

The paper makes three contributions. First, it develops a multidimensional empirical framework for observed RWA liquidity that distinguishes between turnover-based activity, participation breadth, and simple active-versus-inactive states. Second, it extends the discussion beyond single-platform real-estate evidence by comparing observed liquidity across a focused sample of tokenized Treasuries, gold-backed commodities, and private-credit-related assets. Third, it proposes a feasible empirical strategy for a young and data-constrained market, combining descriptive analysis, non-parametric group tests, and exploratory panel regressions that emphasize disciplined association rather than unsupported causal claims.

The remainder of the paper proceeds as follows. Section 2 positions the study within the literature on tokenization, liquidity, and market structure. Section 3 describes the data, defines the liquidity measures and explanatory variables, and presents the empirical strategy. Section 4 discusses the descriptive patterns, group-difference tests, regression results, and broader structural interpretation. Section 5 concludes with implications, limitations, and directions for future research.

\section{Literature Review}

\subsection{Tokenization and the liquidity narrative}

The mainstream case for tokenization is that programmable ledgers can lower frictions across the asset life cycle. WEF emphasizes transparency, efficiency, accessibility, fractional ownership, and composability as core differentiators of tokenized markets \cite{WEF2025}. BIS similarly frames tokenization as the recording of traditional assets on programmable platforms that can combine issuance, trading, settlement, and conditional execution in new ways \cite{BIS2024}. IMF approaches the same issue through the lens of market inefficiencies, asking whether tokenization reduces search costs, transaction costs, information asymmetries, and counterparty frictions across issuance, exchange, servicing, and redemption \cite{IMF2025}.

These potential benefits create a recurring expectation that tokenization should improve liquidity for assets that have historically been hard to trade. OECD explicitly notes that tokenization may improve the tradability of assets with near-absent liquidity, but also warns that liquidity outcomes depend on ecosystem development, interoperability, and the avoidance of market fragmentation \cite{Nassr2025}. Federal Reserve research likewise notes that tokenization may lower barriers to entry and improve liquidity for otherwise inaccessible markets, while also creating interconnections and fragilities if reference assets are themselves illiquid \cite{Carapella2023}.

The central implication for this paper is straightforward. Tokenization can change market design, but the existence of a token does not in itself demonstrate the presence of liquid secondary markets. Liquidity must be measured directly.

\subsection{Empirical evidence from tokenized real estate and related RWAs}

Empirical work on realized liquidity remains limited and concentrated in real estate tokenization. Swinkels documents fragmented ownership but only modest secondary activity among tokenized residential properties, with ownership changing about once per year on average and somewhat more frequently for tokens listed on decentralized exchanges \cite{Swinkels2023}. Kreppmeier et al.\ show that tokenization can broaden access and ownership while leaving investors under-diversified and exposing secondary activity to both property-specific and crypto-market determinants \cite{Kreppmeier2023}. Laschinger et al.\ move closer to the present paper’s concern by focusing directly on liquidity mechanisms in tokenized real estate and highlighting the role of venue design, observable valuations, and market architecture \cite{Laschinger2024}.

More recent evidence suggests that liquidity gains may be conditional rather than general. Cornelli studies U.S.\ tokenized real estate from 2019 to 2025 and finds that tokenization tends to emerge where traditional market liquidity is weaker and access to credit is more limited. He also finds that tokenized real estate can preserve liquidity during exogenous shocks, but only when platforms provide buyback features that create a separate solvency tradeoff \cite{Cornelli2025}. This matters because it suggests that observed liquidity in RWAs is institutionally mediated and may depend on contractual backstops rather than pure market depth.

Taken together, these studies establish two stylized facts. First, tokenization can broaden ownership and lower entry barriers. Second, realized liquidity remains modest, heterogeneous, and highly dependent on venue design and institutional arrangements. The present paper builds on those insights by moving from an asset-class-specific literature to a broader cross-asset measurement framework.

\subsection{Market structure, regulation, and the measurement gap}

Institutional and regulatory analyses consistently emphasize that tokenized liquidity is shaped by market structure. OECD argues that the absence of liquid markets and the absence of a supporting ecosystem can become mutually reinforcing, with issuers hesitant to tokenize without investors and investors hesitant to participate without liquid markets \cite{Nassr2025}. IOSCO similarly observes that native securities have thus far exhibited low secondary-market liquidity because much industry experimentation has focused on infrastructure, primary issuance, settlement, and repo rather than active trading \cite{IOSCO2025}. FSB highlights broader vulnerabilities related to liquidity mismatch, leverage, interconnectedness, and operational design in DLT-based tokenization \cite{FSB2024}.

This leads to an important measurement issue. In traditional finance, liquidity is usually assessed through several dimensions, including trading volume, turnover, bid-ask spreads, depth, and price impact. Public RWA datasets do not yet provide that full menu for most assets. Many tokens have no continuously observed traded price, some trade in permissioned environments, and some only permit access to accredited or whitelisted investors. The practical implication is that RWA liquidity must be measured using observable proxies while acknowledging their limitations.

This paper responds to that measurement gap by defining observed liquidity through token-level platform data and on-chain activity measures, then linking these measures to explanatory variables that are directly available in public RWA datasets.

\section{Data and Methodology}

\subsection{Data sources and sample construction}

The empirical analysis combines two levels of evidence. First, aggregate market information from RWA.xyz is used to describe the broader growth and composition of the tokenized real-world asset (RWA) market. Second, the formal empirical analysis is conducted on a token-month panel constructed from manually collected monthly observations for selected Ethereum-based RWA tokens. The Ethereum restriction is intentional. It provides a more consistent and comparable sample because Ethereum offers the most transparent and verifiable token-level activity data for this set of assets, while also reducing cross-chain measurement inconsistency. Supplemental contract-level observations from Etherscan~\cite{Etherscan} are used only as a supporting reference when token activity patterns require additional verification or illustration.

The empirical unit is the token-month. The analysis covers the six-month period from December 2025 to May 2026 and includes nine non-stablecoin tokenized RWAs: BUIDL, BENJI, OUSG, USTB, USDY, SCOPE, STAC, PAXG, and XAUT. These tokens were chosen to represent three prominent segments of the tokenized RWA market that can be observed consistently in public blockchain data: U.S. Treasury-backed tokens, gold-backed commodity tokens, and private-credit-related tokens. Focusing on these categories makes it possible to compare observed liquidity across asset classes that differ in economic function, participation breadth, and market design. The resulting panel contains 54 token-month observations. USDC was collected separately as a benchmark for scale and market activity, but it is excluded from the baseline estimations because its core economic role is transactional settlement rather than tokenized investment exposure.

For each token-month observation, the dataset records total on-chain asset value, number of holders, monthly transfer volume, and monthly active addresses. The panel is constructed from monthly snapshots taken at calendar month-end or, where an exact month-end value was unavailable, the nearest available end-of-month observation. Using a consistent month-end convention improves comparability across months and supports a balanced descriptive and exploratory panel framework.

An important measurement caveat is that on-chain transfers are not equivalent to economic trades. Monthly transfer volume may reflect genuine secondary-market activity, but it may also include minting and redemption events, treasury movements, custodial rebalancing, or other operational flows. For this reason, the analysis treats transfer-based variables as proxies for observed liquidity rather than direct measures of execution quality, bid-ask spreads, market depth, or price impact. The empirical results should therefore be interpreted as evidence on relative on-chain activity and tradability across tokens and over time, not as a complete market microstructure assessment.

\begin{table}[H]
\caption{Core variables in the empirical design}
\label{tab:variables}
\centering
\small
\begin{tabularx}{\textwidth}{lYY}
\toprule
Variable & Definition & Source and treatment \\
\midrule
Turnover$_{it}$ & Monthly transfer volume divided by total asset value. Measures observed on-chain activity relative to token size. & Constructed from the collected token-month panel using monthly transfer volume and total asset value. Log turnover is used in baseline regressions where transfer volume is strictly positive. \\
\addlinespace
Log active addresses$_{it}$ & Natural logarithm of monthly active addresses. Captures the breadth of monthly participation. & Constructed from the collected monthly active-address observations. Observations with zero active addresses are excluded from log specifications. \\
\addlinespace
Active month$_{it}$ & Indicator equal to one when monthly transfer volume is above zero and zero otherwise. Captures whether any observed monthly activity exists. & Constructed from the collected monthly transfer-volume observations. \\
\addlinespace
Log size$_{it}$ & Natural logarithm of total asset value. Controls for token scale. & Constructed from the collected monthly total-asset-value observations. \\
\addlinespace
Log holders$_{it}$ & Natural logarithm of holders. Captures the breadth of token ownership. & Constructed from the collected monthly holder observations. \\
\addlinespace
Log active ratio$_{it}$ & Natural logarithm of active addresses divided by holders. Measures participation intensity relative to the ownership base. & Constructed as an alternative liquidity proxy in robustness analysis. Defined only when both active addresses and holders are strictly positive. \\
\addlinespace
Asset class fixed effects & Controls for category-specific differences across Treasuries, gold-backed commodities, and private-credit-related tokens. & Constructed from the asset classification assigned to each token in the dataset. \\
\addlinespace
Month fixed effects & Controls for common monthly shocks affecting all tokens in the sample. & Constructed from the monthly panel structure covering December 2025 to May 2026. \\
\addlinespace
Token fixed effects & Controls for time-invariant token-specific heterogeneity in robustness specifications. & Constructed from token identifiers in the dataset. \\
\bottomrule
\end{tabularx}
\end{table}

\subsection{Liquidity measures and explanatory variables}

The baseline analysis employs a size-adjusted proxy for observed on-chain liquidity constructed from the token-month variables in the dataset. Following the standard turnover logic of scaling observed activity by asset size, baseline turnover is defined as:

\begin{equation}
\text{Turnover}_{it} = \frac{\text{Monthly Transfer Volume}_{it}}{\text{Total Asset Value}_{it}}.
\label{eq:turnover}
\end{equation}

This variable captures observed on-chain transfer activity relative to token size. It does not represent a perfect measure of secondary-market liquidity, because transfer volume may include operational or administrative flows in addition to trading. Nevertheless, it provides a more comparable liquidity proxy than raw transfer volume alone, especially when tokens differ sharply in scale.

Given the variables available in the dataset that we collected, the empirical analysis uses two supplementary liquidity indicators. The first is \(\log(\text{Monthly Active Addresses}_{it})\), which captures the breadth of active participation in a given month. The second is \(\text{ActiveMonth}_{it}\), a binary indicator equal to 1 if monthly transfer volume is greater than zero and 0 otherwise. In the robustness analysis, the paper also considers an alternative participation-based measure, the active-address ratio, defined as monthly active addresses divided by total holders.

The explanatory variables are restricted to those consistently observed in the collected panel. The baseline specification therefore includes token size, measured as total asset value; holder breadth, measured as the number of holders; and asset-class indicators for Treasury, Gold, and Private Credit. Month fixed effects are added to capture common temporal shocks affecting all tokens in the sample. In alternative specifications, token fixed effects are also used to absorb time-invariant token-specific heterogeneity.


Table \ref{tab:variables} summarizes the core variables used in the empirical analysis. The baseline liquidity proxy is turnover, defined as monthly transfer volume divided by total asset value, which scales observed activity by token size. In addition to turnover, the analysis uses log active addresses and an active-month indicator to capture alternative dimensions of observed liquidity. The robustness analysis also considers the active ratio, which relates active participation to the broader holder base. All continuous size and participation variables are log-transformed because the data is highly skewed, with substantial differences between small and large tokens.

\subsection{Hypotheses and empirical strategy}

Given the variables available in the token-month panel, the empirical analysis tests three hypotheses.

\begin{enumerate}
    \item \textbf{H1:} Observed liquidity differs across tokenized RWA asset classes.
    \item \textbf{H2:} Tokens with broader holder bases exhibit higher observed liquidity.
    \item \textbf{H3:} Larger tokens do not necessarily exhibit higher observed liquidity once participation breadth and asset-class differences are taken into account.
\end{enumerate}

The empirical strategy proceeds in three stages.

First, the paper reports descriptive statistics and category-level comparisons for the token-month panel based on our dataset. Because the liquidity proxies are skewed and the sample includes very small and very large tokens, medians are emphasized alongside means. The descriptive analysis is used to document the extent of heterogeneity across Treasuries, gold-backed tokens, and private-credit-related assets.

Second, the paper uses non-parametric and rank-based methods to evaluate cross-sectional differences and monotonic relationships. Kruskal--Wallis tests are used to compare liquidity proxies across asset classes because the sample is small and the variables are highly non-normal. Spearman rank correlations are also reported to assess whether broader holder participation is associated with higher observed liquidity. 

Third, the paper estimates exploratory panel-style regressions with month fixed effects to examine the token characteristics associated with observed liquidity. The baseline specification is:

\begin{equation}
\text{Liquidity}_{it} = \alpha + \beta_1 \log(\text{Size}_{it}) + \beta_2 \log(\text{Holders}_{it}) + \gamma_a + \tau_t + \varepsilon_{it},
\label{eq:baseline}
\end{equation}

where \(i\) indexes tokens and \(t\) indexes months. \(\text{Liquidity}_{it}\) denotes the liquidity proxy for token \(i\) in month \(t\). In the main specification, this dependent variable is \(\log(\text{Turnover}_{it})\), where turnover is defined as monthly transfer volume divided by total asset value and is observed only when transfer volume is strictly positive. \(\alpha\) is the constant term. \(\log(\text{Size}_{it})\) is the natural logarithm of total asset value and captures token scale, while \(\log(\text{Holders}_{it})\) is the natural logarithm of the number of holders and captures ownership breadth. \(\gamma_a\) represents asset-class fixed effects, which absorb systematic differences across token categories, and \(\tau_t\) represents month fixed effects, which absorb common monthly shocks. \(\varepsilon_{it}\) denotes the residual error term.

As an additional specification, the analysis also uses \(\text{ActiveMonth}_{it}\) as the dependent variable, where \(\text{ActiveMonth}_{it}=1\) if monthly transfer volume is positive and \(\text{ActiveMonth}_{it}=0\) otherwise. This specification captures the presence of observed monthly activity rather than its turnover intensity. 

\section{Results and Discussion}

\begin{figure}[H]
\centering

\includegraphics[width=0.48\textwidth]{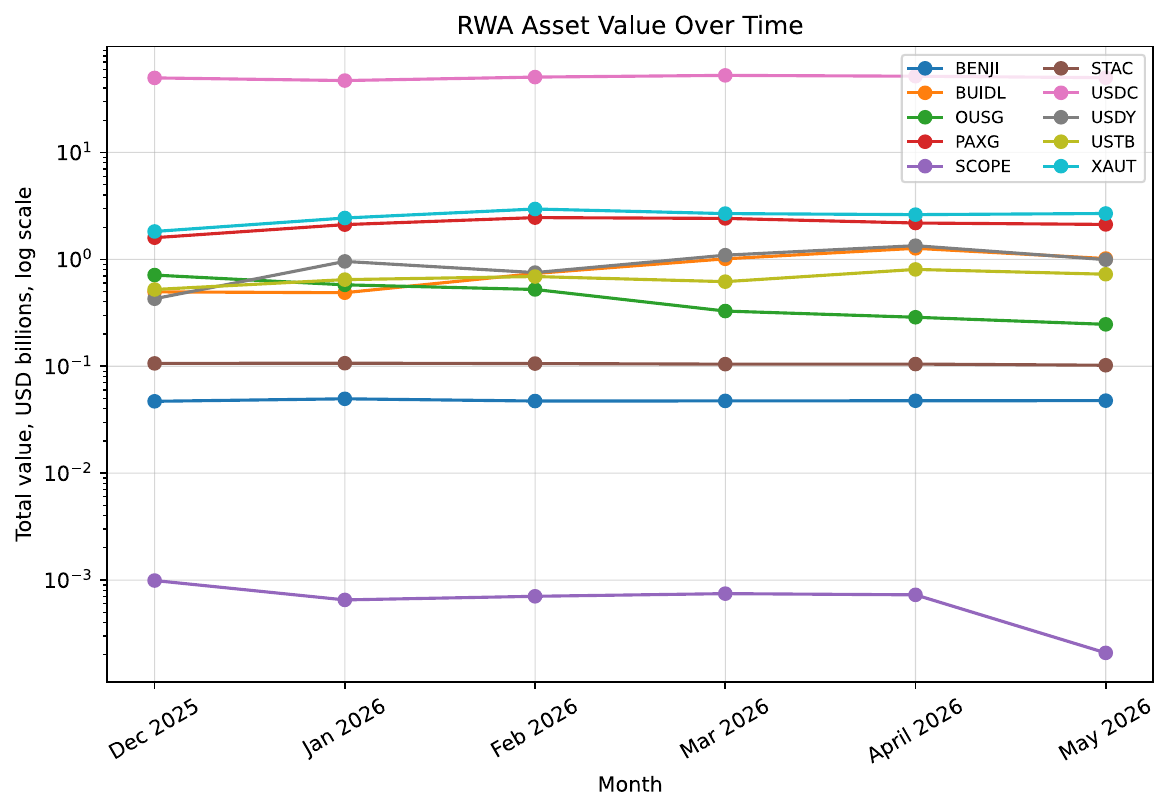}
\hfill
\includegraphics[width=0.48\textwidth]{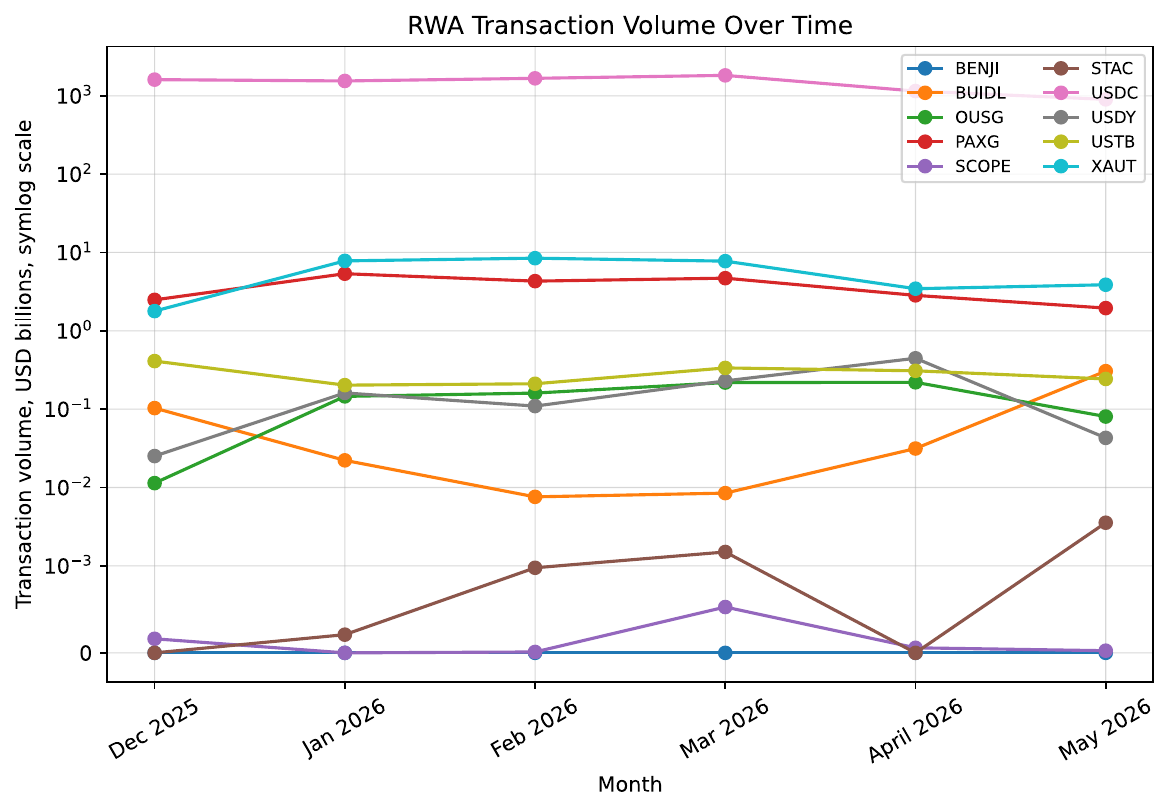}

\caption{Monthly asset value and transaction activity of selected tokenized real-world assets from December 2025 to May 2026. The left panel shows total asset value using a logarithmic scale, while the right panel shows transaction volume using a symmetric logarithmic scale. These figures show that market size and transaction activity do not necessarily move together. Some assets have large asset values but relatively limited transaction volume, while others display stronger on-chain activity relative to their market size.}
\label{fig:value-volume-comparison}
\end{figure}

\begin{figure}[H]
\centering

\includegraphics[width=0.48\textwidth]{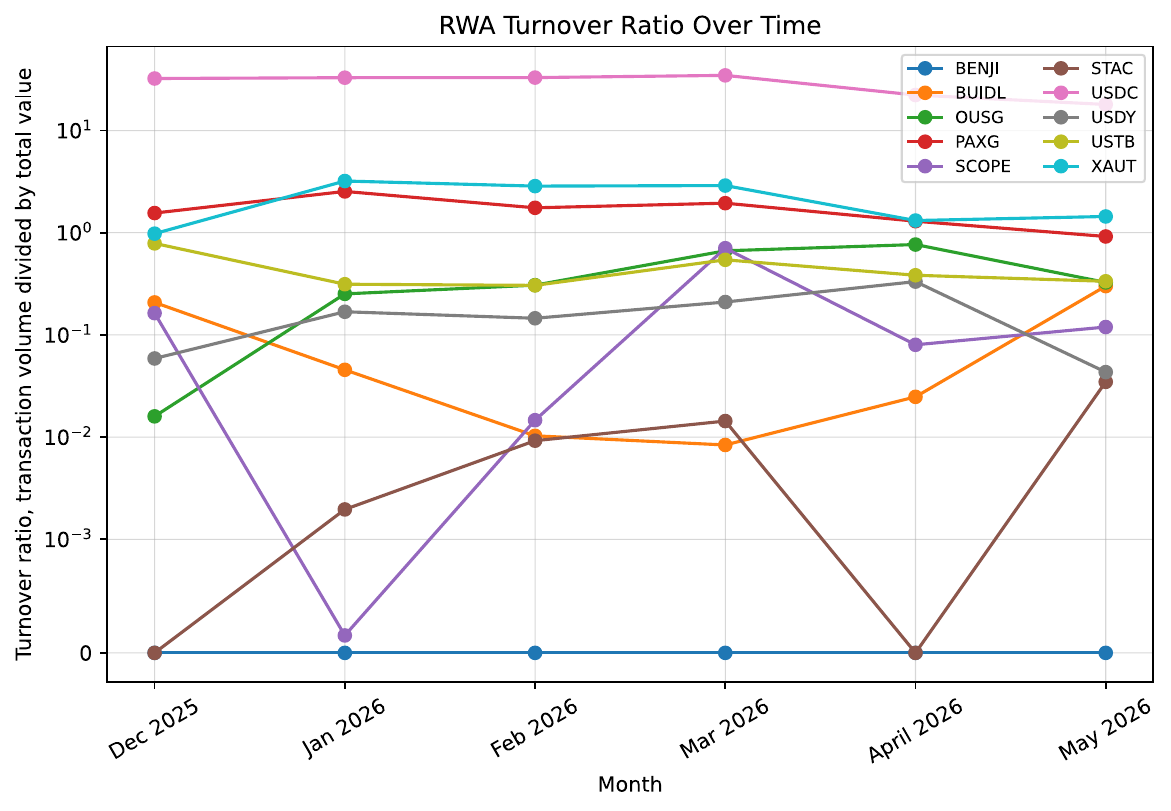}
\hfill
\includegraphics[width=0.48\textwidth]{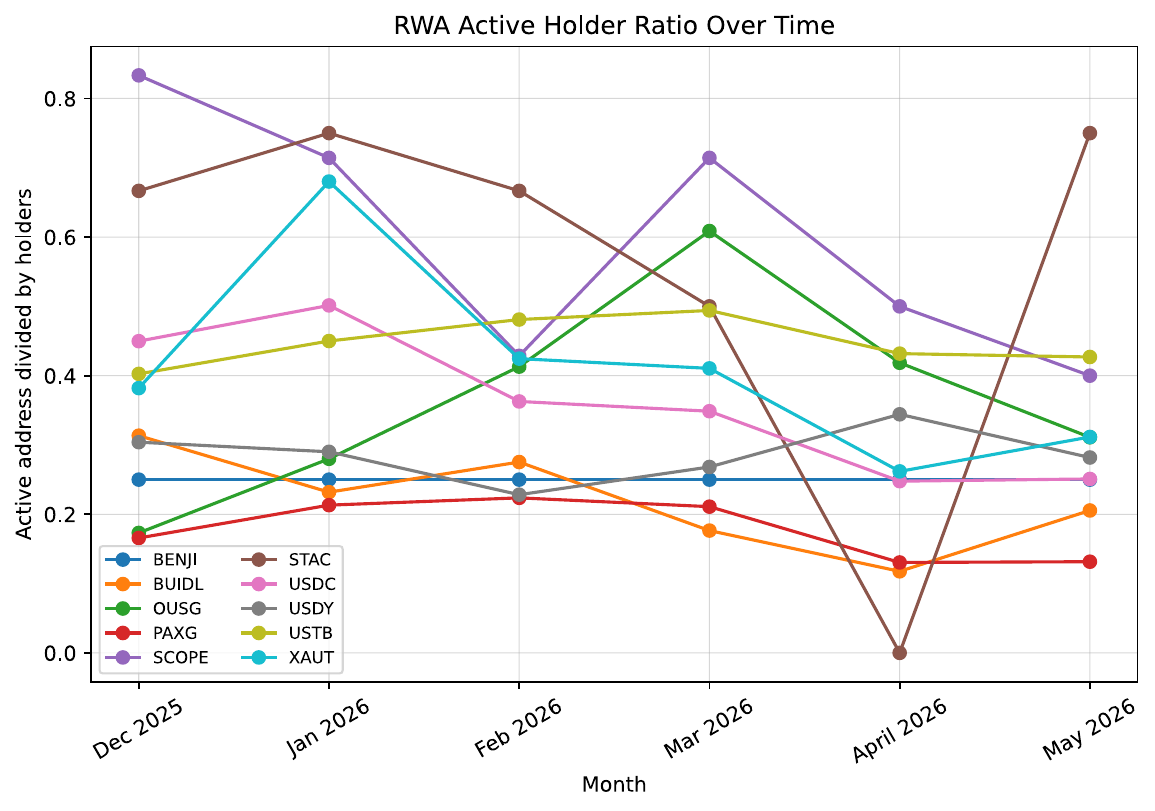}

\caption{Liquidity intensity and active participation of selected tokenized real-world assets from December 2025 to May 2026. The left panel reports the turnover ratio, calculated as transaction volume divided by total asset value. The right panel reports the active holder ratio, calculated as active addresses divided by total holders. Together, these measures provide a more informative view of market quality than total value alone because they capture whether assets are actively transferred and whether holder participation is broad or passive.}
\label{fig:turnover-activeholder-comparison}
\end{figure}

\begin{figure}[H]
\centering

\includegraphics[width=0.48\textwidth]{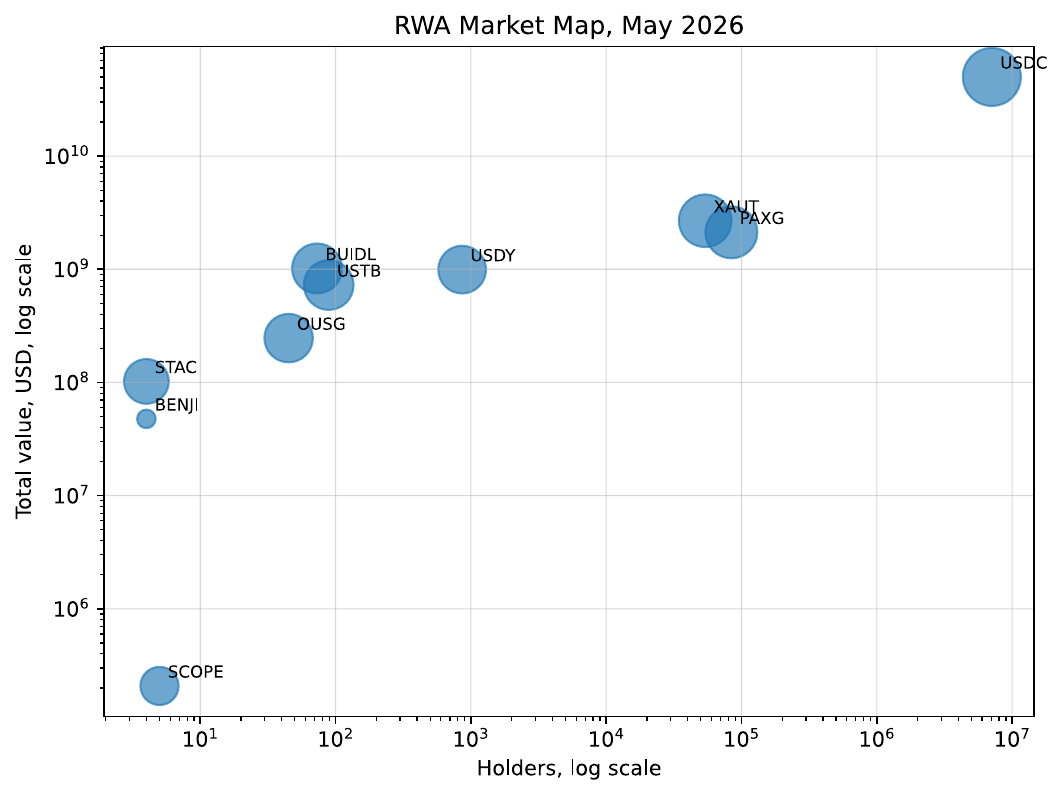}
\hfill
\includegraphics[width=0.48\textwidth]{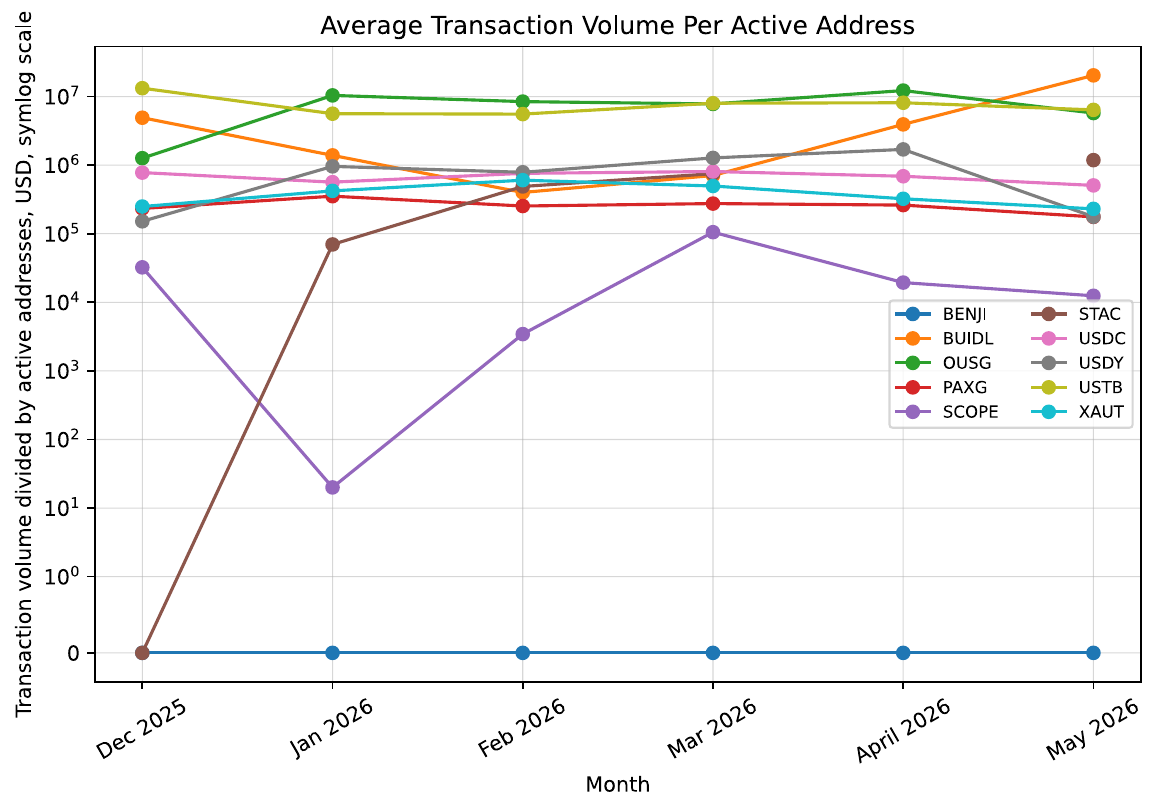}

\caption{Cross-sectional market structure and transaction concentration of selected tokenized real-world assets. The left panel presents the May 2026 market map, where the x-axis shows the number of holders, the y-axis shows total asset value, and bubble size represents transaction volume. The right panel reports average transaction volume per active address, calculated as transaction volume divided by active addresses. These figures help identify whether on-chain activity is broadly distributed across users or concentrated among a smaller number of active addresses.}
\label{fig:marketmap-transactionconcentration}
\end{figure}

The descriptive results are summarised in Figures~\ref{fig:value-volume-comparison}, \ref{fig:turnover-activeholder-comparison}, and \ref{fig:marketmap-transactionconcentration}, which jointly compare market size, transaction activity, liquidity intensity, active participation, and transaction concentration across the selected tokenized real-world assets. Figure~\ref{fig:value-volume-comparison} shows that total asset value varies substantially across assets, with USDC serving as a large stablecoin benchmark, while several tokenized RWA products remain much smaller in scale. However, the comparison between total value and transaction volume indicates that a larger asset base does not necessarily correspond to stronger on-chain activity. Figure~\ref{fig:turnover-activeholder-comparison} further supports this interpretation by showing that liquidity intensity and active participation differ considerably across assets, with some products exhibiting relatively low turnover or passive holder bases despite meaningful asset values. Finally, Figure~\ref{fig:marketmap-transactionconcentration} highlights cross-sectional differences in market structure and transaction concentration, showing whether activity is broadly distributed across users or concentrated among a smaller number of active addresses.

\subsection{Descriptive evidence from the token-month panel}

The token-month panel reveals substantial heterogeneity across the sampled RWA tokens. As shown in Figure~\ref{fig:value-volume-comparison} (right panel) and Figure~\ref{fig:turnover-activeholder-comparison} (left panel), gold-backed tokens such as PAXG and XAUT combine broader holder bases with stronger recurring on-chain activity than most Treasury and private-credit-related tokens. By contrast, BENJI and STAC display near-zero turnover in several months despite maintaining non-trivial asset value, while BUIDL combines relatively large scale with modest participation breadth and uneven activity intensity. These patterns are consistent with a market structure in which some institutional-grade Treasury and credit products function primarily as issuance, yield, or treasury-management instruments rather than actively traded assets.

Table~\ref{tab:desc} summarizes the distribution of the core empirical variables. The results show wide dispersion across the panel. Log turnover has a mean of $-1.667$, a median of $-1.184$, and the largest relative spread, ranging from $-8.782$ to $1.165$, which indicates that observed liquidity varies sharply across token-month observations. Log active addresses is also highly dispersed, with a mean of $3.994$, a median of $2.944$, and a range from $0.000$ to $9.827$, suggesting large differences in participation breadth across tokens and months. The variation in log size, which ranges from $12.247$ to $21.807$, and in log holders, which ranges from $1.099$ to $11.340$, further shows that the sample includes both very small and very large assets with very different ownership profiles.

Taken together, these descriptive patterns support the central argument of the paper. Tokenization does not automatically translate into active secondary-market liquidity. Instead, the panel shows that liquidity conditions differ markedly across token types, with gold-backed tokens occupying the strongest observed position, Treasury tokens displaying an intermediate but internally diverse profile, and private-credit-related tokens remaining the weakest overall.

\begin{table}[H]
\caption{Descriptive statistics of the token-month panel based on available variables}
\label{tab:desc}
\centering
\small
\begin{tabular}{lrrrrr}
\toprule
Variable & Mean & Median & Std.\ Dev. & Min & Max \\
\midrule
Log turnover         & -1.667 & -1.184 & 2.128 & -8.782 & 1.165 \\
Log active addresses &  3.994 &  2.944 & 3.353 &  0.000 & 9.827 \\
Log size             & 19.312 & 20.208 & 2.481 & 12.247 & 21.807 \\
Log holders          &  5.019 &  4.227 & 3.526 &  1.099 & 11.340 \\
\bottomrule
\end{tabular}
\end{table}
\begin{flushleft}
\footnotesize
Notes: All variables are expressed in natural logarithms. Log turnover is computed only for observations with strictly positive transfer volume. Log active addresses excludes observations with zero active addresses.
\end{flushleft}


\subsection{Group differences in observed liquidity}

Table~\ref{tab:grouptests} reports the baseline non-parametric tests of observed liquidity across asset classes. The results show strong evidence of cross-category heterogeneity. The Kruskal--Wallis tests reject equality across asset classes for both log turnover and log active addresses at the 1\% level. Substantively, this finding is consistent with the descriptive pattern in which gold-backed tokens, especially PAXG and XAUT, display materially stronger observed liquidity than Treasury and private-credit-related assets. Gold-backed tokens combine broader holder participation with greater recurring on-chain activity, whereas several Treasury and credit-linked tokens exhibit lower participation intensity and weaker turnover despite sometimes large outstanding asset values.

\begin{table}[H]
\caption{Kruskal--Wallis tests of observed liquidity across asset classes}
\label{tab:grouptests}
\centering
\small
\begin{tabular}{lccc}
\toprule
Comparison & Liquidity proxy & Test statistic & p-value \\
\midrule
Across asset classes & Log turnover & 29.572 & $<0.001$ \\
Across asset classes & Log active addresses & 32.524 & $<0.001$ \\
\bottomrule
\end{tabular}
\end{table}

\begin{flushleft}
\footnotesize
Notes: The Kruskal--Wallis test is used because the liquidity variables are highly skewed and the data is relatively small. Observations with undefined logarithms, such as zero turnover or zero active addresses, are excluded from the relevant test.
\end{flushleft}

\subsection{Panel evidence on liquidity determinants}

Table \ref{tab:panel} reports baseline panel regressions using the available token-month data from the Ethereum-based RWA data. Model (1) uses log turnover as the main proxy for observed trading liquidity, while Model (2) uses an active-month indicator equal to one when transfer volume is positive. Both models include asset-class and month fixed effects in order to account for broad category differences and common monthly shocks.

The results show that holder breadth is positively associated with the probability that a token records non-zero activity in a given month. In Model (2), the coefficient on log holders is positive and statistically significant at the 1\% level, suggesting that tokens with broader participation are more likely to remain active over time. By contrast, log size is not statistically significant in either specification, indicating that larger asset value alone does not guarantee stronger observed liquidity once category and month effects are taken into account.

Relative to the omitted Gold category, both Treasury and Private Credit tokens exhibit significantly lower turnover in Model (1), which is consistent with the descriptive evidence that gold-backed tokens such as PAXG and XAUT display stronger on-chain liquidity profiles. In Model (2), the positive coefficients on Treasury and Private Credit should be interpreted cautiously, as the binary active-month variable captures any non-zero transfer activity rather than the intensity of trading. Taken together, the regressions suggest that participation breadth matters more than sheer size, while category differences remain central to observed liquidity outcomes.

\begin{table}[H]
\caption{Baseline panel regressions of observed liquidity}
\label{tab:panel}
\centering
\small
\begin{tabular}{lcc}
\toprule
 & (1) Log turnover & (2) Active month \\
\midrule
Log size & -0.309 & -0.004 \\
         & (0.206) & (0.034) \\
Log holders & -0.121 & 0.202*** \\
            & (0.216) & (0.038) \\
Private credit & -7.400*** & 1.681*** \\
               & (2.010) & (0.332) \\
Treasury & -3.554*** & 1.157*** \\
         & (1.225) & (0.242) \\
Asset-class fixed effects & Yes & Yes \\
Month fixed effects & Yes & Yes \\
Observations & 46 & 54 \\
Tokens & 8 & 9 \\
$R^2$ & 0.602 & 0.563 \\
\bottomrule
\end{tabular}
\begin{flushleft}
\footnotesize
Notes: Heteroskedasticity-robust standard errors are reported in parentheses. The omitted asset class is Gold. Model (2) is estimated as a linear probability model, where ActiveMonth equals 1 if monthly transfer volume is positive and 0 otherwise. Log turnover is only defined for observations with strictly positive transfer volume, so zero-turnover months are excluded from Model (1). *** $p<0.01$.
\end{flushleft}
\end{table}

\begin{table}[H]
\caption{Robustness checks using alternative sample restrictions and specifications}
\label{tab:robust}
\centering
\small
\begin{tabular}{lccc}
\toprule
 & (1) Token fixed effects & (2) Excluding gold-backed tokens & (3) Log active ratio \\
\midrule
Log size & -0.169 & -0.296 & 0.035 \\
         & (0.871) & (0.197) & (0.028) \\
Log holders & -2.164 & -0.096 & -0.016 \\
            & (2.035) & (0.262) & (0.028) \\
Treasury & Absorbed by token FE & 3.669*** & 0.076 \\
         &                      & (1.021) & (0.236) \\
Private credit & Absorbed by token FE & Reference & 0.896*** \\
               &                      &           & (0.284) \\
Month fixed effects & Yes & Yes & Yes \\
Token fixed effects & Yes & No & No \\
Observations & 46 & 34 & 53 \\
Tokens & 8 & 6 & 9 \\
\bottomrule
\end{tabular}
\begin{flushleft}
\footnotesize
Notes: Heteroskedasticity-robust standard errors are reported in parentheses. Column (1) replaces asset-class fixed effects with token fixed effects, so asset-class coefficients are absorbed and not separately identified. Column (2) excludes PAXG and XAUT to test whether the baseline findings are driven by the unusually liquid gold-backed tokens. In Column (2), Private Credit is the omitted asset class. Column (3) uses log active ratio, defined as the natural logarithm of active addresses divided by holders, as an alternative proxy for observed liquidity. In Column (3), Gold is the omitted asset class. *** $p<0.01$.
\end{flushleft}
\end{table}

Table~\ref{tab:robust} reports three robustness checks based on alternative specifications and sample restrictions. Column~(1) replaces asset-class fixed effects with token fixed effects in the log-turnover regression. In this specification, the coefficients on log size and log holders remain statistically insignificant, suggesting that the baseline turnover results are not driven solely by stable cross-token differences. Because token fixed effects absorb all time-invariant token characteristics, asset-class coefficients are not separately identified in this column.

Column~(2) excludes the two gold-backed tokens, PAXG and XAUT, which are the most liquid assets in the sample. In this restricted specification, the Treasury coefficient remains positive and statistically significant relative to the omitted Private Credit category. This indicates that Treasury tokens continue to exhibit stronger turnover than thinner credit-linked assets even after the unusually liquid gold-backed tokens are removed from the sample.

Column~(3) uses log active ratio, defined as the natural logarithm of active addresses divided by holders, as an alternative proxy for observed liquidity. In this specification, the coefficient on Private Credit is positive and statistically significant relative to the omitted Gold category, while the Treasury coefficient is small and statistically insignificant. This result suggests that participation intensity relative to the holder base captures a different dimension of market activity than turnover. In other words, a token category may appear weak in turnover terms while still showing comparatively high activity relative to its ownership base. Taken together, the robustness checks suggest that the main conclusions are reasonably stable, although the short sample and limited covariate set mean that all robustness results should still be interpreted with caution.

\subsection{Interpretation of the main empirical findings}

The empirical results provide support for all three hypotheses, although the strength of the evidence differs across them.

First, the results support \textbf{H1}, which states that observed liquidity differs across tokenized RWA asset classes. The descriptive analysis, the Kruskal--Wallis tests in Table~\ref{tab:grouptests}, and the baseline regressions in Table~\ref{tab:panel} all point in the same direction. Gold-backed tokens, particularly PAXG and XAUT, occupy the strongest observed liquidity position in the sample. They combine broader holder bases, larger active-address counts, and higher turnover than most Treasury and private-credit-related tokens. By contrast, several Treasury and private-credit-related assets remain much thinner despite sometimes substantial asset value. This result indicates that token category is not merely a background classification, but an economically meaningful dimension of liquidity outcomes.

Second, the results provide support for \textbf{H2}, which states that tokens with broader holder bases exhibit higher observed liquidity. In the baseline panel regressions, log holders is positive and statistically significant in the active-month specification, indicating that tokens with broader participation are more likely to record non-zero monthly activity. Although the holder variable is not statistically significant in every turnover-based specification, the overall pattern still suggests that participation breadth matters for whether a token remains active over time. This finding is consistent with the descriptive evidence that the most liquid tokens in the sample also tend to have the broadest ownership base.

Third, the results are consistent with \textbf{H3}, which states that larger tokens do not necessarily exhibit higher observed liquidity once participation breadth and asset-class differences are taken into account. Across the baseline and robustness regressions, log size is not statistically significant. This is an important result for the interpretation of tokenized RWA markets because it separates issuance scale from liquidity intensity. A token may have a large outstanding asset value and still display weak turnover or limited recurring activity. In the present sample, size alone is therefore not a reliable indicator of observed liquidity.

Taken together, these findings reinforce the paper's central argument that tokenization and liquidity should not be treated as synonymous. The ability to issue an asset on-chain does not automatically produce active secondary-market use. Instead, the evidence suggests that observed liquidity depends more on participation breadth and asset type than on raw scale alone. In practical terms, this means that the success of tokenization should not be judged only by growth in outstanding value, but also by whether tokens become meaningfully tradable after issuance.

\section{Conclusion}

This paper asks whether tokenized real-world assets actually become more liquid once they are represented on-chain. Based on the descriptive, non-parametric, and exploratory panel evidence from the Ethereum-based token-month panel used in this paper, the answer is no in any automatic sense.

The paper contributes by operationalizing observed RWA liquidity through publicly observable on-chain proxies and by linking those proxies to token characteristics. The central empirical insight is that large outstanding asset value does not, by itself, demonstrate liquid secondary markets. In the observed dataset from December 2025 to May 2026, gold-backed tokens such as PAXG and XAUT display the strongest observed liquidity, combining broader holder participation with higher turnover and more persistent activity. By contrast, several Treasury and private-credit-related tokens exhibit weaker and more uneven liquidity despite sometimes substantial asset value. The results therefore support the view that participation breadth and asset category matter more than raw scale alone for observed liquidity outcomes.

This study has several limitations. The dataset is restricted to nine Ethereum-based non-stablecoin RWA tokens, with USDC retained only as a benchmark, observed over a relatively short six-month period. In addition, the available dataset is limited to four core variables: total asset value, holder count, transfer volume, and active addresses. As a result, the analysis cannot directly model several factors that are likely to matter for liquidity, including transfer restrictions, redeemability, ownership concentration, and venue access.

These limitations also define the next research agenda. Longer token histories would support stronger panel designs and allow more credible analysis of liquidity persistence and changes over time. Broader token coverage across multiple chains would help determine whether the patterns documented here generalize beyond the Ethereum-based sample. Richer data on transferability rules, redemption rights, ownership concentration, and trading venues would also enable a more complete modeling of liquidity determinants. Even so, the central conclusion remains clear: tokenization changes the form of ownership, but liquid secondary markets require additional conditions related to participation breadth, accessibility, and market design that cannot be assumed to emerge automatically once an asset is placed on-chain.

\end{document}